\documentclass[aps,pre]{revtex4}
\usepackage{amssymb}
\usepackage{amsmath}
\usepackage{graphicx}
\usepackage{float}

\begin{document}

\title{Behaviour of circular chains of nonlinear oscillators with Kuramoto-like local coupling.}
\author{K. \surname{Garc\'ia Medina}}
\author{E. \surname{Estevez-Rams}}
\email{estevez@fisica.uh.cu}
\affiliation{Facultad de F\'isica-Instituto de Ciencias y Tecnolog\'ia de Materiales(IMRE), Universidad de La Habana, San Lazaro y L. CP 10400. La Habana. Cuba.}

\begin{abstract}
The conditions under which synchronization is achieved for a one-dimensional ring of identical phase oscillators with Kuramoto-like local coupling are studied. The system is approached in the weakly coupled approximation as phase units. Instead of global couplings, nearest-neighbor interaction is assumed. Units are pairwise coupled by a Kuramoto term driven by their phase differences. The system exhibits a rich set of behaviors depending on the balance between the natural frequency of isolated units and the self-feedback. The case of two oscillators is solved analytically, while a numerical approach is used for $N>2$. Building from Kuramoto, the approach to synchronization, when possible, is studied through a local complex order parameter. The system can eventually evolve as a set of coupled local communities towards a given phase value. However, the approach to the stationary state shows a non-monotonous non-trivial dynamic.
\end{abstract}


\date{\today}
\maketitle
   
    \section{Introduction}
Synchronization is ubiquitous in many systems, from physics to social \cite{strogatz12}. Many different processes and actions in nature require precise timing, and, in some specific examples, this timing is so exact that it seems to be the result of a long rehearsal. As an example, consider the composing cells of a healthy beating heart, all in perfect synchrony; or neurons in a brain when reading and interpreting reality as understood in the binding problem \cite{Singer_I,Singer_II,Farmer}. In other systems, synchronization appears spontaneously after a transient period. In society, the claps at the end of a play where every person in the audience starts clapping at their own independent frequency, to end up clapping in synchrony with the entire auditorium; or the well known case of synchronous flashing in South Asian fireflies \cite{buck1968mechanism}. 

Research on synchronization phenomena focuses mainly on identifying and understanding the mechanisms responsible for synchronous behavior. One of the first mathematical studies on collective synchronization was that by Wiener \cite{Wiener_I,Wiener_II}, who speculated about the relation between synchronous activity and alpha rhythms in the brain. Winfree \cite{Winfree} formulated the problem in terms of a large population of limit-cycle oscillators. There are several possible synchronous behaviors in these systems, so it is common to talk in terms of phase synchronization or coherence and frequency synchronization or phase-locking. Winfree noticed that for nearly identical weakly coupled units, the system could be characterized using only their phases so that the model could be understood in terms of a phase motion equation for each oscillator. A further simplification of the model is a mean-field approximation where each unit is coupled to the collective rhythm created by the population. The coupling is taken to be phase-dependent. Winfree \cite{Winfree} found a phase transition in the population of oscillators. Whenever the deviation in frequency distribution was significant compared to the coupling, the system would behave incoherently. However, a small phase-locked group of oscillators would appear if such deviation is reduced.

Kuramoto \cite{Kuramoto_I} gave a strong confirmation to Winfree's intuition by obtaining a universal phase equation for the long-term dynamics of any system of weakly coupled, nearly identical limit-cycle oscillators. Even after reduction to a phase model, the Kuramoto equations are extremely difficult to analyze since the coupling and the system topology is not fixed. The most tractable case is a mean-field approximation. Thus, Kuramoto's work focused on the simplest possible case of equally weighted, all-to-all sinusoidal interaction, depending on the phase differences. Positive feedback between coupling and coherence is then found. The more coherent the population becomes, the more effective the coupling is, being able to recruit even more oscillators to the coherent group. The mechanism behind spontaneous synchronization, intuitively identified by Winfree, stands out clearly in the Kuramoto model \cite{Kuramoto_II}. This model has been used to study synchronization between oscillatory units, being simple enough to render analytical solutions but general enough to display complex behavior and a wealth of synchronization phenomena \cite{Acebron,Strogatz}.

Less is known, however, for locally coupled systems \cite{Kopell,Yanchuk}. There is still the question about how coordinate or complex behavior can emerge from the activity of an individual, locally interacting units \cite{Winfree}. In many cases, the dynamics of isolated units are well known, yet the question about the mechanisms behind collective behavior remains unanswered. 

Recently, Alonso proposed a nearest-neighbor coupling of Adler oscillators. Despite its simple coupling, it shows the emergence of complex collective behavior with the tuning of the control parameters of the system \cite{ALZ}. It has been discussed \cite{ALZ,Magnasco} that such locally balanced interaction poises the system in a massive high-dimensional Hopf bifurcation enabling long-range correlations in a network of excitable units. Surprisingly, such a simple graph topology can give rise to a rich set of behaviors. It has been assumed that collective behavior can only happen with non-local couplings in more involved topologies in coupled oscillators. However, Alonso's report and further works pointed to a different picture. Alonso \cite{ALZ} was able to identify three different kinds of spatiotemporal patterns and classify them as absorbing for quickly decaying solutions; complex for solutions that triggered long complicated transients that may not decay to periodic attractors; and chaotic for solutions that quickly evolve into disordered states with no evident regularities. Further discussions about the phenomenology of this model were studied by Estevez et al. \cite{Estevez}. It was even reported that in the boundary between the chaotic and complex region, enhanced computation at the edge of chaos could be verified \cite{estevez19}. 

As interesting as Alonso's toy model may be, the chosen local coupling is not taken as the more justified Kuramoto coupling, even preserving the nearest neighbor topology. If this last coupling is assumed, will the system still exhibit the emergence of complex dynamics? In this contribution, we start studying such a system. The emergence of complexity will not tackle directly, but instead, an analysis of the stability and collective correlation between oscillators in the control parameter space will be studied. This will start laying the necessary background to, in a second article, study the emergence of complexity and its dependence on the coupling type. 

A ring of nearest neighbors interacting oscillator is used as a model. Each unit is independently excited, and the locally coupled system is left to evolve while looking for the emergence of collective behavior. As the focus is on the interplay between local coupling and global synchronization, the exact nature of the oscillators is not of a big concern, and the proposed model builds from coupled Adler oscillators, where each unit evolves according to \cite{Adler}

\begin{equation}\label{eq:adler}
\dot{\theta}_i=\omega+\gamma \cos(\theta_i).
\end{equation}

where $\gamma$ determines the strength of the self-feedback, and $\omega$ is the natural frequency of the unit.

The model studied in this contribution was taken to explore the possibility of synchronization of the nonlinear Adler oscillators, not achieved in the Alonso model, and to evaluate the influence of the specific coupling form in the long-term dynamics of the system. For this purpose, a set of $N$ locally coupled oscillatory units is considered using a Kuramoto-like local coupling. Each identical unit interacts only with its nearest neighbors in a way that excitation and inhibition are locally balanced. Weak phase coupling is assumed. The state of each oscillatory unit is determined by its phase $\theta_i$, and the corresponding equations of motion are 

\begin{equation}\label{newmodel}
\dot{\theta}_i=\omega+\gamma \cos(\theta_i)+(-1)^i p \left[ \sin(\theta_{i-1}-\theta_i)+\sin(\theta_{i+1}-\theta_i)\right],
\end{equation} 

where $i \in \left[1,N\right]$, $\omega$ and $\gamma$ can be interpreted as in the Adler equation, and $p$ stands for the coupling strength between the units. Notice how the coupling term is proportional to the sine of the phase differences between neighbors, unlike the one proposed on \cite{ALZ}, which depended on the cosines of the phases of the neighboring units. This means that, in the proposed model, the farther the neighbors are on the phase circle (up to $\pi/2$), the stronger they interact. In what follows, and for the rest of the article, $p$ will be taken as $1$, and the results are given in units related to $p$. Note that the alternating sign on the coupling term, given by $(-1)^i$ guarantees the local balance of the interactions. Oscillatory units are either purely excitatory or inhibitory and are arranged in a one-dimensional network where periodic boundary conditions are imposed. Finally, $N$ is taken as an even number, so local coupling balance is achieved everywhere.

The paper is organized as follows. Section \ref{sec:n2} deals with the simplest case of two coupled oscillators (N=2). Equilibrium points on the phase space are discussed, as well as sufficient conditions for coherence to appear.  Section \ref{sec:n4} deals with the $N>2$ case. Possible mechanisms behind the observed synchronization behaviors are discussed in Section \ref{ri}. Finally, the conclusions follow.

\section{Two coupled (N=2) oscillators.}\label{sec:n2}

There is no simple analytic treatment to the system described by equation (\ref{newmodel}), being a many-body nonlinear problem, yet, important insight can be gained by studying specific cases. The simplest case is the two coupled oscillators (N=2) where the phase space becomes a $2$-torus, and the evolution of the system is governed by
\begin{equation}
\begin{array}{l}
\dot{\theta}_1=\omega+\gamma \cos(\theta_1)-2 \sin(\theta_2-\theta_1)\\\\
\dot{\theta}_2=\omega+\gamma \cos(\theta_2)-2 \sin(\theta_2-\theta_1).
\end{array}\label{eq:n2}
\end{equation}

For $\omega$, $\gamma> 0$ equilibria ($\dot{\theta}_1=\dot{\theta}_2=0$) is achieved at 
\begin{equation}
cos(\theta_1^*)=\cos(\theta_2^*)=-\frac{2}{\gamma}\sin(\theta_1^*-\theta_2^*)-\frac{\omega}{\gamma}=a,\label{eq:equlibria}
\end{equation}
with solutions of the form
\begin{equation}{}
\begin{array}{l}
\theta_1^*= \pm \arccos\left[a\right]\\\\
\theta_2^*= \pm \arccos\left[a\right].
\end{array}\label{eq:sols}
\end{equation}
which exist for $\vert a \vert \leq 1$. 

Equilibria points will appear in two forms: ($\theta^*$, $\theta^*$) and ($\theta^*$, $-\theta^* $). For the former, it holds that
\begin{equation}
\cos\theta^*=-\frac{\omega}{\gamma}.\label{eq:equaltheta}
\end{equation}
and the condition for the existence of equilibria points in the form  ($\theta^*$,$\theta^*$) is then,
\begin{equation}
\omega\leq \gamma. \label{eq:fcond}
\end{equation}

When $\omega=\gamma$ a single equilibria point is observed at $\theta^*=\pm \arccos\left[-1\right]=\pm \pi$. While for $\omega < \gamma$, a pair of equilibria points emerges, one for each $\pm \arccos\left[a\right]$ branch.\\

For equilibria points in the form ($\theta^*$,$-\theta^*$), equation (\ref{eq:equlibria}) can be reduced to the quadratic form
\begin{equation}
x \left (\gamma\pm4\sqrt{1-x^2}\right)+\omega=0, \label{eq:fcond1}
\end{equation}
where $x=\cos \theta^*$. 

\begin{table}[!ht]
{\centering
\begin{tabular}{lc}
Region & No. eq. points \\\hline
I      &  0   \\
II     &  2   \\
III    &  4   \\
IV     &  6   \\
V      &  4   \\
\hline
\end{tabular}
}
\caption{Number of equilibria points in the different regions of configuration space for two coupled oscillators (N=2). The number of equilibria points can go from none in the region I, up to  $6$, in region IV.}\label{tbl:equilibria}
\end{table}

\begin{figure}[ht!]%
\begin{center}
\includegraphics[keepaspectratio=true,width=.9\textwidth]{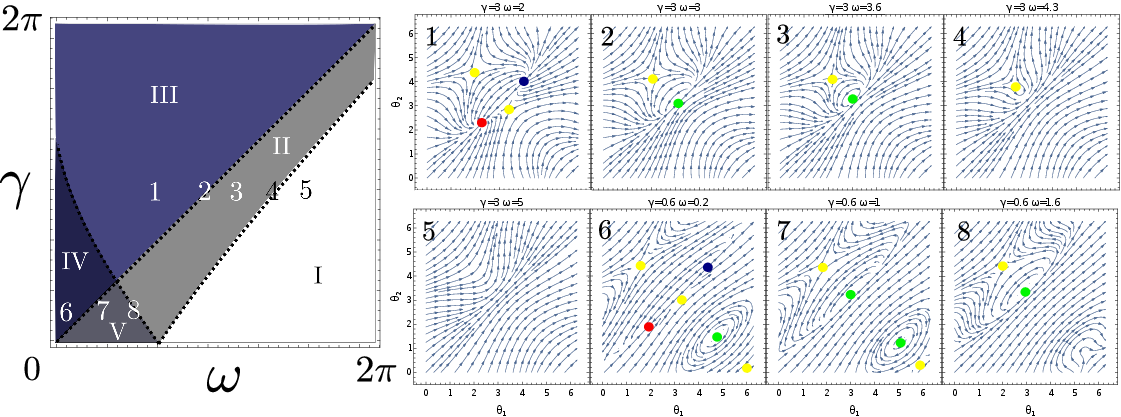}
\caption{\textbf{Left} Schematic representation of the different regions in parameter space, according to the number of equilibria solutions. \textbf{Right} Vectorial flow diagrams along the lines $\gamma=3$ (points from $5$ to $1$) and $\gamma=0.6$ (points from $8$ to $6$) for $N=2$. Arrows indicate the direction of the flow. Equilibria points appear as stable nodes (red), unstable nodes (blue), saddle points (yellow), and nonlinear centers (green). The system undergoes several bifurcations as $\omega$ is varied, showing up to six different equilibria points in region IV.}\label{fig:equilibria}
\end{center}
\end{figure}

Combination of Equations (\ref{eq:equaltheta}) and (\ref{eq:fcond1}) results in five regions according to the number of possible solutions (Figure \ref{fig:equilibria}). The number of equilibria points in each region is summarized in table \ref{tbl:equilibria}.

In region I, there are no equilibrium points. Nevertheless, a slow bottleneck region is observed on the central portion of the flow diagram portrait (see Fig. \ref{fig:equilibria} diagram 5). At the boundary between region I and II (Fig. \ref{fig:equilibria} diagram 4), the slowing-down and bending of trajectories on the bottleneck become more evident, and a saddle point (yellow) appears with attracting trajectories on its stable manifold and repealing ones, on its unstable manifold. When crossing the boundary towards region II (Fig. \ref{fig:equilibria} diagram 3), a second equilibrium point emerges as a result of a first bifurcation. A new equilibrium point (green) is found as a nonlinear center. The regions around this equilibria point determine closed orbits circling it. This implies that for any pair of initial ($\theta_{1}$,$\theta_2$) near enough to this point, the system gets trapped indefinitely in such cyclic orbits which do not get closer, nor farther to the steady point. The point is neither attractive nor repulsive, and the system would have periodic behavior around it.

Within region III  (Fig. \ref{fig:equilibria} diagram 1), the saddle point previously observed exhibits no qualitative change. However, the nonlinear center at ($\pi$,$\pi$) that appeared at the boundary (Fig. \ref{fig:equilibria} diagram 2) splits into three different equilibria points. One of those points is stable (red), corresponding to the positive branch of $\arccos[\omega/\gamma]$, and another unstable point (blue) corresponding to the negative branch also emerges. Lastly, a second saddle point appears, corresponding to one of the solutions to equation (\ref{eq:fcond1})  of the form $\theta_2^*=-\theta_1^*$. In region III, the single stable point is globally attractive, as confirmed by the computation of Lyapunov exponents. Figure \ref{fig:liapunov} shows the largest Liapunov exponent ($\lambda$) dependence on $\omega$ for $\gamma=3$ (see supplementary materials for computation details). Notice how the diagonal $\gamma=\omega$ corresponds to a sign shift on $\lambda$, indicating a qualitative change in the system dynamics. For any initial ($\theta_{1}$,$\theta_2$) values, the system ends up settling at the stable configuration ($\arccos[\omega/\gamma]$,$\arccos[\omega/\gamma]$).

\begin{figure}[ht!]%
\begin{center}
\includegraphics[scale=0.6]{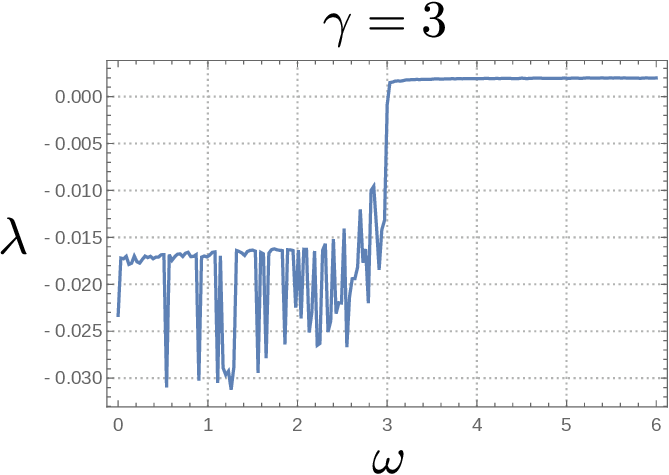}	
\caption{ The behavior of the largest Liapunov exponent ($\lambda_M$) with parameter $\omega$, for $N=2$ and $\gamma=3$. The jump in the Lyapunov exponents confirms the observed bifurcation when crossing the diagonal $\gamma=\omega$. (The presented plot is the average over $n$ different initial conditions for each $\omega$ value. Fluctuations were found to decrease with increasing $n$. The figure corresponds to $n=500$.)}\label{fig:liapunov}
\end{center}
\end{figure}

For lower $\gamma$ value, in the boundary between region II and V (Fig. \ref{fig:equilibria} diagram 8), only two equilibrium solutions are reported, namely, the saddle point and the nonlinear center already observed for region II. However, a previously unobserved bottleneck region appears in the lower-right corner of the configuration portrait, mimicking the one that gave rise to the mentioned saddle and center points. The same kind of bifurcation appears here since a new well-defined saddle point and a nonlinear center is observed in region V (Fig. \ref{fig:equilibria} diagram 7). The system in region V has two saddle points and two nonlinear centers, all aligned along the diagonal $\theta_2=-\theta_{1}$. This means the system can now be captured by two different periodic behaviors, even though no attractors are yet observed. 

Finally, in region IV, steady solutions are available ($\theta_1^*$,$\theta_1^*$). As a result, one of the two nonlinear centers splits into a third saddle point, a stable and an unstable node. In this region of the parameter space, the system displays six equilibria points, three saddle points, a stable node, an unstable node, and a nonlinear center. The stable node is not globally attractive. Closed trajectories, like the ones surrounding nonlinear centers, are characteristic of conservative and reversible systems. However, the presence of a stable node in the configuration portraits indicates that our system is not conservative. Furthermore, the model is not reversible either. Notice how equations (\ref{newmodel}) are sensitive to the changes of $t \rightarrow -t$ and $\theta_{i} \rightarrow -\theta_{i}$ $\forall i \in [1,N]$. \\

In summary, borders between different regions in Figure \ref{fig:equilibria} are the bifurcation lines for this system. Only two kinds of bifurcation occur. Hamiltonian saddle-node bifurcations (saddle-center) \cite{garcia2020tilting, diminnie2000slow, dullin2005another} take place when crossing from the region I to II, II to V, and III to IV, creating a bi-dimensional saddle point and a nonlinear-center, each time. On the other hand, what appears to be a multi-dimensional saddle-node bifurcation takes place when crossing the diagonal line, creating two nodes of opposite stability and a multi-dimensional saddle point.

\section{$N>2$ coupled oscillators.}\label{sec:n4}

For $N>2$, a similar analysis can be performed. Equilibria configurations are achieved when $\dot{\theta_i}=0$  $\forall i \in [1,N]$. This implies, using equation (\ref{newmodel}), that 
\begin{equation}
\gamma \cos(\theta_i^*)=(-1)^{i+1} \kappa_i -\omega\label{eq:SecondCondition}
\end{equation}
with 
\begin{equation}\label{eq:lambda}
\kappa_i=\left\lbrace \sin(\theta_{i-1}-\theta_i)+\sin(\theta_{i+1}-\theta_i)\right\rbrace
\end{equation}

Determination of all possible equilibria configurations is a rather complicated task since the actual steady phase distribution will strongly depend on the neighboring phase's relation at a local scale throughout $\kappa_i$. However, important information about the system can be drawn by noticing that whenever $\theta_{i-1}=\theta_{i}=\theta_{i+1}$ equation (\ref{eq:lambda}) yields $\kappa_i=0$. If the previous is true for all sites in the system, then equation (\ref{eq:SecondCondition}) results in 
\begin{equation}
\cos(\theta_i^*)=-\omega/\gamma\;\;\;\forall i \in [1,N] 
\end{equation}
This means that a globally coherent state where 
\begin{equation}
\theta_i^*= \pm \arccos[-\omega/\gamma]\;\;\;\forall i \in [1,N]\label{eq:equi1}
\end{equation}
is always a possible equilibrium state with existence condition $\omega \leq \gamma$, regardless of the system's size. It is easy to demonstrate, by direct substitution, that the condition $\pm \theta_{i-1}=\mp \theta_{i}=\pm \theta_{i+1}$ also renders a steady solution to equations (\ref{newmodel}), with a steady phase value given by

\begin{equation}
\gamma \cos(\theta_i^*)=2 (-1)^{i} \sin(2 \theta_i^*) -\omega
\end{equation}

These sign combinations correspond to alternating patterns similar to ones reported by Alonso \cite{alonso17} on the absorbing regime.  

\begin{figure}[ht!]%
\begin{center}
\includegraphics[keepaspectratio=true,width=.5\textwidth]{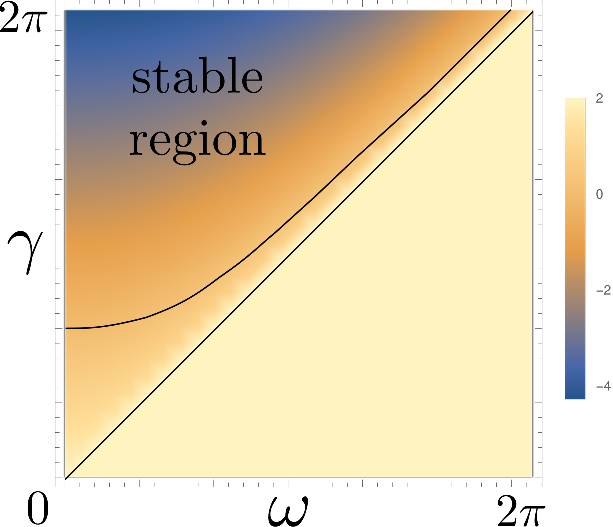}	
\caption{Behavior of the real part of the largest Lyapunov exponent, $Re(\lambda_M)$, on parameter space for the steady state given by (\ref{eq:equi1}). Stability is limited to a specific region. There is a portion of parameter space (between both solid lines) where the coherent steady solution exists, yet is unstable.}\label{fig:Stability_Phase_Locked}
\end{center}
\end{figure}

Linear stability of a given equilibria configuration can be determined by looking at the real part of the biggest eigenvalue $\lambda_M$ of the Jacobian matrix of the system  (\ref{newmodel}) evaluated on the equilibrium configuration. Stability was numerically explored for the coherent state described by equation (\ref{eq:equi1}) for all even $N$ in the interval $\left[4,500\right]$. 

Figure \ref{fig:Stability_Phase_Locked} shows the behavior of $Re(\lambda_M)$ for the state corresponding to the positive branch of equation (\ref{eq:equi1}). The stability region, where $Re(\lambda_M)<0$, showed no significant dependence on the system size. Notice how the bifurcation curve for this type of solution ($\omega=\gamma$) does not coincide with a sign shift on $Re(\lambda_M)$, which means that stability depends on particular values of the control parameters. This is a new feature compared to the $N=2$ case, where $\theta_{1}=\theta_2=\arccos\left[-\omega/\gamma\right]$ is always stable beyond bifurcation. 

Proof of stability for any even $N$ is yet to be found.\\

\section{Local complex order parameter }\label{ri}

In order to study the mechanisms behind the appearance of a globally coherent state, building from Kuramotos' \cite{Kuramoto_I,Kuramoto_II} approach, a local, complex order parameter is defined. For a system where identical oscillatory units are connected to their $2d$ closest neighbors, such a local order parameter could be defined as 
\begin{equation}\label{localOrder}
r_{k}e^{i\Phi_k}=\frac{1}{1+2 d}\sum_{j=k-d}^{k+d}e^{i\theta_{j}}
\end{equation}
Notice that if the units in the group are oscillating with similar phases, $r_k$ is equal to $1$ and $\Phi_k=\theta_k$. On the other hand, whenever the units have very different phases, $r_k$ will tend to zero as $d\rightarrow \infty$ (for $d=1$, as in the model described by equation (\ref{newmodel}), the mean value of $r_k$ with random phases is $0.525$). In this sense, $r_k$ is a measure of order and synchrony at a local scale. The smaller its value, the less synchronized the phases are. Equation (\ref{newmodel}) can be rewritten in terms of the new parameters. For this purpose, equation (\ref{localOrder}) is multiplied by $e^{-i \theta_{k}}$ and the imaginary parts of both members are equated in the resulting expression, thus obtaining
\begin{equation}
r_k (2d+1) \sin(\Phi_k-\theta_{k})=\sum_{j=k-d}^{k+d}\sin(\theta_{j}-\theta_{k}),\label{eq:rk}
\end{equation} 
using equation (\ref{eq:rk}), expression (\ref{newmodel}) becomes
\begin{equation}
\dot{\theta}_k=\omega+\gamma \cos(\theta_k)+(-1)^k r_k (2d+1) \sin(\Phi_k-\theta_{k}).\label{eq:ne}
\end{equation} 

Equation (\ref{eq:ne}) makes explicit that the coupling term depends on the number of connected neighbors $2d$. This means that a fully connected infinite population, generalization of our model, would require normalizing the coupling term with respect to the number of units in the system ($N$) to avoid the divergence of the interaction strength, as in the original Kuramoto model \cite{Kuramoto_I}. 

In this formulation, phases appear to evolve independently from each other, but the interaction is modeled through $r_k$ and $\Phi_k$. This can be considered a local mean-field approximation, where phases interact with the local average phase $\Phi_k$ rather than with each other. An effective coupling strength depending on the degree of local synchrony, is in place. The coupling strength term is now proportional to $r_k$, creating positive feedback between coupling and coherence. The more coherent the population at a local scale, the strongest the coupling becomes for units in that group.

This explains how locally synchronized behavior would appear. However, more is needed to answer the question of how synchrony goes from a local to a global scale. The answer to this question is competition. Notice how $r_k$ and $r_{k \pm 1}$ are not independent. They overlap on their definition given by equation (\ref{localOrder}) since different communities or groups have common units. These communities compete with each other as they try to recruit units. Each community evolves to a different locally-coherent state. However, since they share individuals, global equilibrium is only achieved when all communities are in the same locally coherent state, which turns out to be globally coherent.

\begin{figure}[ht!]%
\begin{center}
\includegraphics[keepaspectratio=true,width=.5\textwidth]{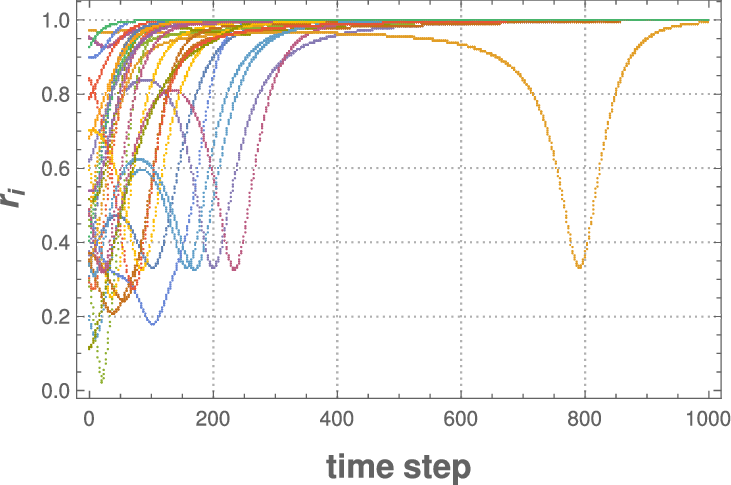}	
\caption{Evolution over time of $r_i$ for 30 randomly selected units. Changes in the monotony of $r_i$ point to the existence of competition among neighboring communities with different degrees of coherence can be seen as a possible mechanism for coherence propagation along the system. The simulation was performed for $N=500$, $\omega=3$, and $\gamma=4$, starting from a random configuration and evolving for a thousand time steps with $\Delta t=0.005$. }\label{fig:Order_Parameter}
\end{center}
\end{figure}

Figure \ref{fig:Order_Parameter} shows the evolution of $r_k$ over time, for $30$ randomly selected units, in a $500$ units population. Simulations were performed for $\omega=3$ and $\gamma=4$, starting from a random set of phase values uniformly distributed in [0, 2$\pi$] and evolving for a thousand times steps. A Runge-Kutta O(4-5) routine was used with a maximum error of $\delta=10^{-6}$ while solutions were sampled with a $\Delta t=5\times 10^{-3}$ step. The different curves in the figure seem to confirm the competition hypothesis. For some units, $r_k$ evolves monotonically towards the final maximum value, as should be expected from the positive feedback between coupling and synchrony. However, for a certain group of units, $dr_k/dt$ changes sign with time. This can be understood in terms of competition with neighboring communities. If a given community is more synchronized than another, common units will "prefer" to synchronize to the first one, as the effective coupling strength is higher. This will be reflected in a lowering of both effective coupling and coherence degrees in the second community.

Since all neighboring communities have common units, the fact that $r_k=1$ $\forall k \in [1, N]$ implies global coherence. Nevertheless, important information about how each community evolves towards the globally coherent state can be obtained using the local mean phase $\Phi_k$. 

\begin{figure}[t!!!]%
\begin{center}
\includegraphics[scale=0.4]{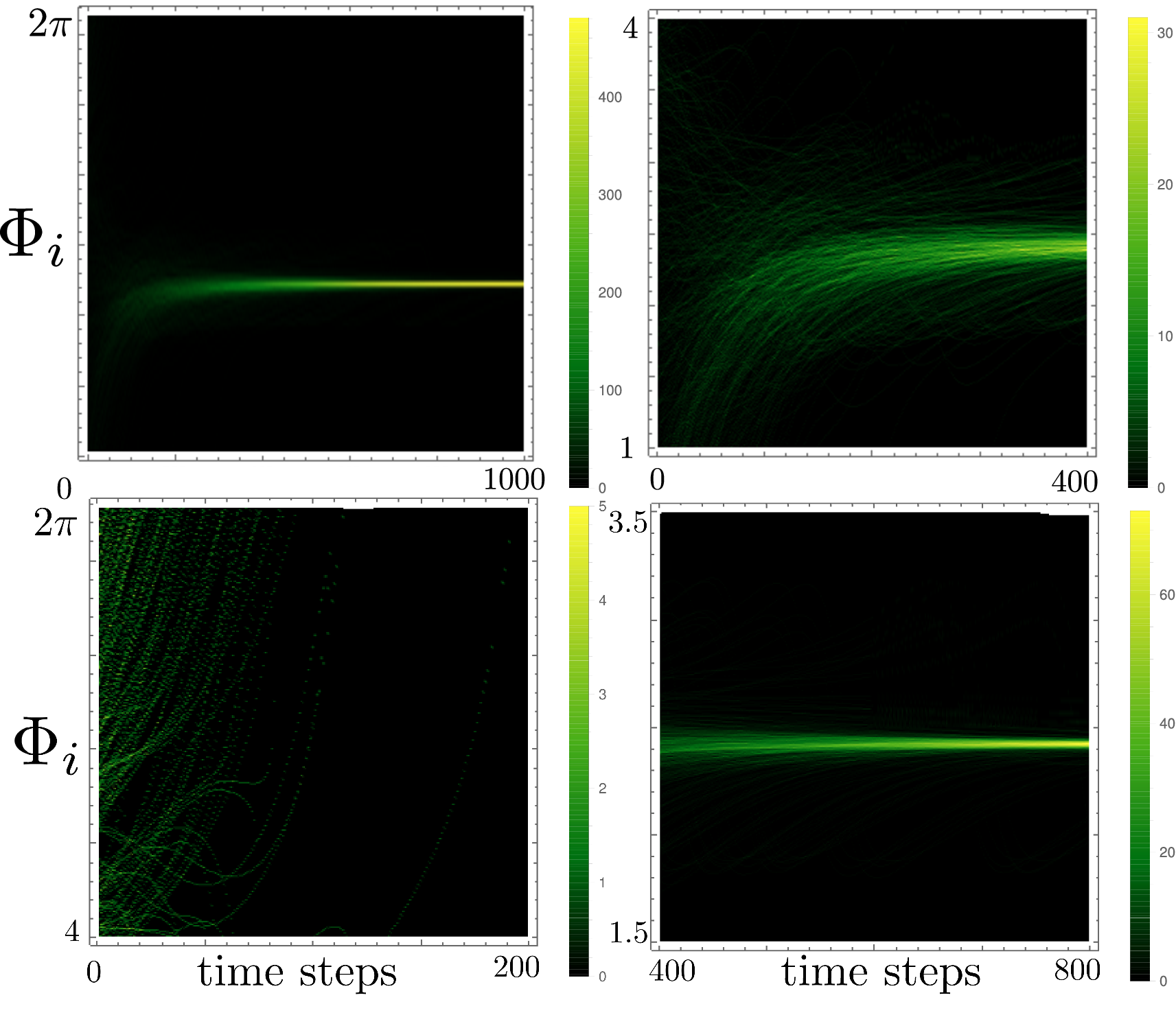}	
\caption{Temporal evolution of the histogram of the community's phase values towards the final ($t\gg1$), steady, globally coherent state with $\Phi_k=\Phi_c=2.42$. Each column corresponds to the histogram at a fixed time.  \textbf{Top left} $\Phi_k$ values range in the entire interval $[0,2\pi]$ and for time steps up to $1000$. Bins were taken with a size of $2 \pi/100$.  \textbf{Top right} Initial evolution for the lower values of $\Phi_k$ up to 400 time steps. Bins are taken with size $3/400$. Communities that start below the convergent value $\Phi_c$ evolve monotonically towards the stationary value. \textbf{Bottom left} Initial evolution up to 200-time steps and for the higher values of $\Phi_k$. Bins are taken of size $(2\pi - 4)/400$. Communities that start above the convergent value $\Phi_c$ can exhibit a more rich evolution towards the stationary value before finally converging to $\Phi_c$. \textbf{Bottom right} Long term evolution, bins are taken with size $(3.5-1.5)/400$. In all cases, simulations were performed for $N=500$, $\omega=3$, and $\gamma=4$.}\label{fig:Mean_phase}
\end{center}
\end{figure}

Figure \ref{fig:Mean_phase} shows the histogram evolution of the $\Phi_k$ value over time. To build this graphic, the interval $[0,2\pi]$ was divided into $100$ sub-intervals (bins), and the system was left to evolve over 1000 time steps. At each instant, the local mean phase $\Phi_k$ was determined at each site, and the number of communities at each sub-interval (bin) was updated. Simulations were performed in the same conditions as before.

Looking at Figure \ref{fig:Mean_phase}, one can see key aspects of the synchronization process in terms of phase values. There is no preferential phase at the beginning since the system starts from a random configuration. As time increases, communities cluster around a poorly defined asymptotic value $\Phi_c=2.42$. For longer times, this asymptotic phase value becomes better defined, and communities are more densely grouped around it. For times over $500$ time steps, global coherence is almost achieved as most communities are grouped close to the same asymptotic phase value, given by $\arccos[-\omega/\gamma]$. Only a few communities remain unsynchronized, and they progressively evolve towards that steady state until, for sufficiently long times, the entire system shares the same $\Phi_c$ and global coherence is achieved.

As already noticed (see \textbf{top right} panel in Figure \ref{fig:Mean_phase}), the system starts with no apparent preferential phase value. However, for low times it can already be seen in complex traffic of communities between different bins or intervals. This means that even at low times, the system quickly evolves towards its attractor at a local scale. The presence of a smooth, clearly upwards flow should be noticed in the graphic going from low local mean phase values to the one corresponding to the steady coherent state, namely $\arccos[-\omega/\gamma]$. This flow manifests as long clear lines starting at phase values below the asymptotic value and grouping around it. This could indicate that those communities that start with $\Phi_i$ values below $\arccos[-\omega/\gamma]$ tend to locally and gradually change into greater $\Phi_i$ values until the asymptotic value is reached. At that point, they tend to stay close to it.

Things change when looking at communities, starting with $\Phi_k$ values above $\arccos[-\omega/\gamma]$. There is a dense pattern of lines right above the value corresponding to the steady coherent state, which can be observed on the \textbf{top right} panel as well, where the general tendency is to follow the grouping behavior around the phase value defined by $\arccos[-\omega/\gamma]$. A second group of lines, however, mainly those starting at $\Phi_k$ values above $4$ (showed on the \textbf{bottom left} panel), go up on the phase scale, getting further away from the asymptotic value, circling the entire range of possible $\Phi_i$ values and taking more time to, eventually, join the above mentioned smooth upwards flow and collapse into $\arccos[-\omega/\gamma]$. In terms of communities, this could indicate that for those that started with $\Phi_k$ values above $\arccos[-\omega/\gamma]$, there are two possible behaviors. The ones that started relatively close to the asymptotic phase value will lower their local phase values and gradually get closer to the asymptotic value. The rest of the communities, which started with $\Phi_k>4$, will tend to evolve towards greater $\Phi_k$, eventually reaching the asymptotic phase value.

As a result of all these processes, the number of communities with $\Phi_k$ around $\arccos[-\omega/\gamma]$ gradually increases until global coherence is achieved. An interesting feature of this synchronization process is what could be called satellite communities. Notice, at the \textbf{top right} panel in Figure \ref{fig:Mean_phase}, the presence of fluctuating lines along the dense cluster around the asymptotic phase value. These lines appear early in the system's evolution. They persist even at times when most communities are already densely grouped around the asymptotic value, as shown in the \textbf{bottom right} panel of the same figure. They are the last and slowest lines to reach $\arccos[-\omega/\gamma]$. These lines could indicate that some specific communities, or groups of them, start fluctuating around the asymptotic phase value while the number of communities with a local mean phase equal to the asymptotic value grows. Only when the vast majority of communities in the system are almost synchronized with each other do these satellite communities gradually decay to a steady, globally coherent state.

\section{Conclusions}\label{Conclusions}

Synchronization in locally coupled excitable oscillatory units with Kuramoto-like coupling was studied in the weak coupling approximation. This can be seen as an instance of Winfree's idea about the role of interaction with collective rhythms, enabling global synchronization, even when local interactions are involved. For $N=2$, the sufficient conditions for phase synchronization to occur in a one-dimensional system of identical units were $\omega < \gamma$.  The complete analysis of critical points for this $N=2$ case was performed. For the $N>2$ case, the region of parameter space where global coherence is stable was also identified. Building from Kuramoto's approach, a local order parameter and a local mean phase value were introduced. The local order parameter can be used to follow the competition among local communities, which is key to understanding the dynamics of global synchronization in locally coupled systems. Simulations showed that not all communities evolve towards global coherence in the same way, giving rise to satellite communities. Further features of these communities, as well as their role in the synchronization process, are yet to be explored. Finally, the fact that global coherence is reachable and stable for only a specific region of parameter space poses the possibility for complex and chaotic behavior in our model, which is why a complexity analysis could capture more interesting features of it and shed more light on the underlying mechanisms behind collective behavior in locally interacting systems.

\section{Acknowledgments}
EER and KGM would like to thank Alexander von Humboldt Stiftung for financial support. EER would like to thank the MPI-PKS for a visiting grant. EER thanks H. Kantz at MPI-PKS for the in-depth discussion and the excellent working environment.

\section{Data Availability}

Data sharing is not applicable to this article as no new data were created or analyzed in this
study.


\end{document}